\documentclass[prb,twocolumn,superscriptaddress]{revtex4}
\usepackage{amssymb, amsmath}

\def\sw{$S(\omega)$}

\def\tlo{0.05 K}
\def\thi{1.6 K}
\def\pr{48.6 bar}
\def\w{$\omega$}
\def\Df{$\Delta$f}
\def\Dq{$\Delta$Q}

\def\gapx{\ \lower 2pt \hbox{$\buildrel>\over{\scriptstyle{\sim}}$}}
\def\lapx{\ \lower 2pt \hbox{$\buildrel<\over{\scriptstyle{\sim}}$}}
\def\lap{\ \lower 2pt \hbox{$\buildrel<\over{\scriptstyle{\sim}}$}}
\def\2d{$2\Delta$}
\def\3he{$^3$He}
\def\4he{$^4$He}
\def\A{\AA$^{-1}$}
\def\a1{$a_{1}$}
\def\a2{$\overline\alpha_{2}$}

\def\Am3{\AA$^{-3}$}
\def\ax2{$a_{2}$}

\def\d2o{D$_2$O}

\def\dr{$\Delta$}

\def\h2o{H$_2$O}

\def\Ke3{$\langle{K_3}\rangle$}

\def\s1{$S_1(Q,E)$}

\def\sq{$S(Q)$}
\def\sqw{$S(Q,\omega)$}

\def\tc{$T_c$}

\usepackage{graphicx}

\begin{document}
\title{Excitations of amorphous solid helium}

\author{Jacques Bossy}                                                                                          
\affiliation{Institut N\'{e}el, CNRS-UJF, BP 166, 38042 Grenoble Cedex 9, France}
\author{Jacques Ollivier}                                                                                         
\affiliation{Institut Laue-Langevin, BP 156, 38042 Grenoble, France}
\author{Helmut Schober}                                                             
\affiliation{Institut Laue-Langevin, BP 156, 38042 Grenoble, France} 
\affiliation{Universit\'{e} Joseph Fourier, UFR de Physique, F38041 Grenoble Cedex 9, France}     
\author{H. R. Glyde}
\affiliation{Department of Physics and Astronomy, University of Delaware, Newark, Delaware 19716-2593, 
USA}

\date{\today}

\begin{abstract}

We present neutron scattering measurements of the dynamic structure factor, \sqw, of amorphous solid 
helium confined in 47 \AA~pore diameter MCM-41 at pressure \pr. At low temperature, $T$ = 0.05 K, we 
observe \sqw~ of the confined quantum amorphous solid plus the bulk polycrystalline solid between the 
MCM-41 powder grains. No liquid-like phonon-roton modes, other sharply defined modes at low energy 
($\omega<$ 1.0 meV) or modes unique to a quantum amorphous solid that might suggest superflow are 
observed. Rather the \sqw~ of confined amorphous and bulk polycrystalline solid appear to be very similar. 
At higher temperature ($T>$ 1 K), the amorphous solid in the MCM-41 pores melts to a liquid which has a 
broad \sqw~ peaked near \w~$\simeq$ 0 characteristic of normal liquid \4he under pressure. 
Expressions for the \sqw~ of amorphous and polycrystalline solid helium are presented and compared. In 
previous measurements of liquid \4he confined in MCM-41 at lower pressure the intensity in the liquid 
roton mode decreases with increasing pressure until the roton vanishes at the solidification pressure (38 
bars), consistent with no roton in the solid observed here.

\end{abstract}
\maketitle

\section{Introduction}

The superfluid fraction of liquid \4he, both bulk and \4he in porous media, is traditionally and most 
accurately measured in a torsional oscillator (TO). Below a critical temperture \tc, the TO frequency 
increases indicating that a fraction of the helium mass has decoupled and ceased to rotate with the TO. 
The effect is denoted a nonclassical rotational inertia (NCRI) and the decoupled fraction is identified as 
the superfluid fraction.   

In 2004, Kim and Chan\cite{Kim:04,Kim:04a} reported a similarly NCRI in solid \4he, in both bulk 
solid\cite{Kim:04} and in solid \4he confined in Vycor.\cite{Kim:04a} Remarkably, below a \tc~$\simeq$ 200 
mK, a small fraction of the solid apparently decouples in a TO. The NCRI has now been widely reproduced in 
other laboratories in a variety of sample 
cells.\cite{Rittner:06,Kim:06,Rittner:07,Kondo:07,Aoki:07,Gumann:11,Fefferman:12}  However, the magnitude 
of the NCRI observed varies between zero and 1.5 \%, depending on how the sample is prepared, quenched or 
annealed, on the \3he concentration and on other factors. This suggests that the NCRI is associated with 
defects in the solid, dislocations, grain boundaries, amorphous regions or other 
defects.\cite{Prokofev:07,Balibar:08,West:09a,Balibar:10} 
The magnitude and character of the NCRI depends on the oscillator frequency\cite{Aoki:07,Gumann:11,Mi:12} 
and shows effects\cite{Hunt:09,Pratt:11a} not usually associated with superflow. For example, the NCRI is 
associated with substantial elastic energy dissipation in the solid described by the Q of the oscillator. 
Indeed in some cases, the observed frequency shift, \Df, and the disipation, \Dq, can be quite well 
described by the real and imaginary parts of a common dynamic susceptibility as found in purely glassy 
systems\cite{Hunt:09,Pratt:11a}.

Day and Beamish\cite{Day:07,Day:09} and others\cite{Rojas:10} have shown that the shear modulus, $\mu$, of 
solid helium increases at low temperature with both a temperature dependence and a dependence on \3he 
concentration that tracks that of the observed NCRI. The increase in $\mu$ is attributed to the stiffening 
of the solid as dislocations become pinned by \3he at low temperature. A key question is whether the \Df~ 
can be entirely attributed to elastic behavior or whether there is some remaining part that must arise 
from other effects such as superflow.

Pursuing this question, Maris\cite{Maris:12}and Beamish and coworkers\cite{Beamish:12} have shown that in 
some TOs there is sufficient solid helium in the torsional rod to explain the observed \Df~ completely in 
terms of elastic stiffening of the solid in the rod. However, in many others it cannot\cite{Beamish:12} 
and at least part of the observed \Df~ must have some other origin. Similarly, Choi and 
coworkers\cite{Choi:10,Choi:12}
using a TO which includes DC rotation have shown that the critical velocity depends on the DC velocity as 
expected for genuine superflow. Thus, while the stiffening of the shear modulus below \tc~ and elastic 
effects can account for the \Df~ and \Dq~ in some cases, it cannot account for \Df~ in all 
cases\cite{Kim:04,Kim:04a,Rittner:08,Choi:10,Fefferman:12} nor all effects.\cite{Kim:11,Choi:10,Choi:12}

Path Integral Monte Carlo calculations predict that the superfluid fraction and Bose-–Einstein condensate 
fraction in perfect crystalline solid helium are negligibly 
small.\cite{Ceperley:04,Boninsegni:06,Clark:06}  However, a finite and observable superfluid 
fraction and condensate fraction is predicted\cite{Boninsegni:06} in amorphous solid helium. The first one 
to two layers of helium on rough porous media walls are amorphous. Typically, the solid in porous media 
grows from the amorphous layers inward toward the center of the pores.\cite{Rossi:05} In Vycor and aerogel 
the tightly bound amorphous layers give way to crystalline solid after a few layers so that the solid in 
the interior of the pore is crystalline. However, if the pore size is small enough, the solid is amorphous 
throughout the pore, as predicted for classical solids.\cite{Coasne:06} Specifically, in 47 \AA~ pore 
diameter MCM-41 and 34 \AA~ pore diameter gelsil, we have shown\cite{Bossy:10} that the entire solid is 
amorphous (no Bragg peaks). Since superflow in amorphous solid \4he is predicted, it is interesting to 
determine whether the amorphous solid has any low lying modes similar to the phonon-roton mode of liquid 
\4he, that might suggest Bose-Einstein condensation (BEC) and superflow, or whether it has vibrational 
modes similar to those of a typical polycrystalline solid as observed in classical amorphous 
solids.\cite{Suzuki:87,Suck:88,Mentese:02}  This is particularly interesting since a low energy mode in 
solid helium in aerogel has recently been reported and identified as the origin of local 
superfluidity.\cite{Lauter:11a}
In this context we present measurements of the dynamical structure factor (DSF) of amorphous solid helium. 

\section{Experiment}   

The experiment was performed at the Institut Laue-Langevin
on the time of flight spectrometer IN5. We used an incident neutron wavelength
of 5 \AA, which provided a spectrometer energy resolution of 85 $\mu$eV.

The sample cell was a cylindrical aluminum container with an inner
diameter of 15 mm and a height of 55 mm. The MCM-41 powder
sample occupied a volume of 7 cm$^3$ corresponding to a height
of 40 mm in the cell. The upper part of the cell containing
bulk He was masked with cadmium.
The cell was mounted in a dilution refrigerator built at
the ILL  that has a base temperature of 40 mK.
The cell has been filled at T=3.8 K and 80 bar, and
the solid sample between the grains of the MCM-41 was formed by the blocked capillary method.

\begin{figure}[h,b,t]
\hspace{0.0cm}
\includegraphics[angle=0,width=0.53\textwidth]{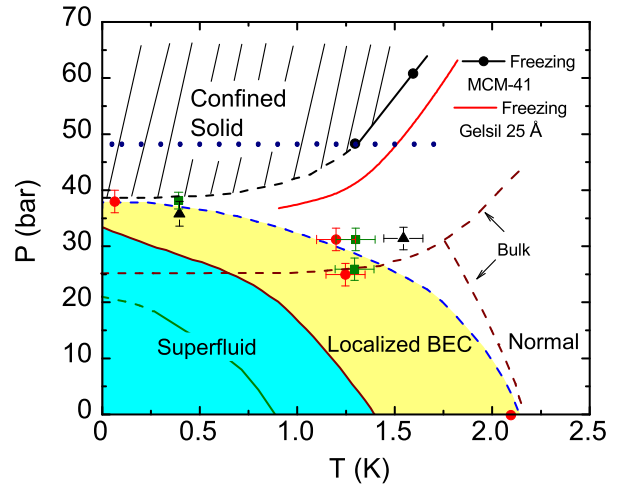}
\caption{Phase diagram of liquid \4he in MCM-41 (47 \AA), gelsil (25 \AA), and FSM (28 \AA). The 
superfluid (SF) phase in FSM and in gelsil is at the lower left, the blue region. The green line is the 
\tc~ of the SF phase in FSM observed by Taniguchi et al.\cite{Taniguchi:10} and  red line the \tc~ in 
gelsil observed by Yamamoto et al.\cite{Yamamoto:04}. The yellow region shows the Localized Bose-Einstein 
condensation (LBEC) region in gelsil. In the LBEC region, well defined phonon-roton modes are 
observed\cite{Pearce:04a,Bossy:08}, up to but not beyond the pressures and temperatures indicated by the 
blue dashed line in gelsil and MCM-41 which is taken as the upper limit of the LBEC region. The red solid 
line is the freezing onset of \4he in gelsil measured by Shirahama et al.\cite{Shirahama:08}. The black 
solid and dashed lines show the freezing onset of \4he in MCM-41  observed\cite{Bossy:08} from the static 
structure factor, \sq. The dashed red lines show the phase boundaries in bulk \4he. The blue dots at $p$ = 
46.8 bar mark the temperatures at which the present measurements were made}
\label{f1}
\end{figure}

In the present 47 \AA~ pore diameter MCM-41 powder sample, the ratio of the volume in the pores, $V_P$, to 
that between the grains, $V_{IG}$, is \cite{Bossy:10} $V_P/V_{IG} = 0.44$. Thus approximately 30 \% of the 
helium in the beam is in the pores, 70 \% between the grains. In a 47 \AA~pore approximately 30\% of the 
volume in the pore is occupied by the tightly bound amorphous solid layers on the media walls. Thus, in 
the present MCM-41 sample at \thi~ when there is liquid in the pores, approximately 80\% of the helium in 
the cell is solid (between the grains and on the pore walls) and only 20\% is liquid (in the pores). For 
this reason, the difference in the scattering intensity arising from the solidification of the liquid to 
an amorphous solid between \thi~and \tlo~is expected to be small, as observed in Figs.~\ref{f2} and 
\ref{f3} below.

\section{Results}

To set the stage, we show the phase diagram of \4he confined in 47 \AA~pore diameter MCM-41 
\cite{Bossy:08,Bossy:10} and in 25 \AA~mean pore diameter gelsil 
\cite{Yamamoto:04,Yamamoto:08,Shirahama:08} in Fig.~\ref{f1}. At low temperature and pressure, liquid \4he 
in 25 \AA~gelsil (and in 28 \AA~FSM ) is a superfluid, as shown by Yamamoto et al. \cite{Yamamoto:04} and 
Taniguchi et al. \cite{Taniguchi:10}, respectively. In the superfluid phase, the associated Bose-Einstein 
condensation (BEC) is expected to be connected and continuous across the sample providing a continuous 
phase that enables macroscopic superflow. At higher temperature above the superfluid phase in porous media 
the liquid forms a localized BEC (LBEC) region in which the BEC is localized to islands. In the LBEC 
region the BEC is isolated in patches, is not extended, and there is no macroscopic superflow across the 
sample \cite{Glyde:00,Plantevin:02}. The LBEC region lies between the superfluid and normal liquid phases
as shown in Fig ~\ref{f1}. In 47 \AA~ MCM-41 and 25 \AA~ gelsil helium solidifies at pressures, $p \gtrsim 
38$ bars. The liquid-solid boundary in MCM-41 has been determined by cooling the liquid and observing a 
small reproducible increase in the peak height of the static structure factor S(Q) on solidification. 
\cite{Bossy:10} In 25 \AA~ gelsil it has been determined from pressure drop on 
solidification.\cite{Shirahama:08} 
No Bragg peaks were observed in the confined solid\cite{Bossy:10} showing that the solid is amorphous with 
no crystalline regions. In larger pore diameter media, for example in 70 \AA~gelsil,\cite{Wallacher:05} 
and aerogel,\cite{Mulders:08} polycrystalline solid \4he regions are observed.

\begin{figure}[h]
\hspace{0.05cm}
\includegraphics[angle=0,width=0.5\textwidth]{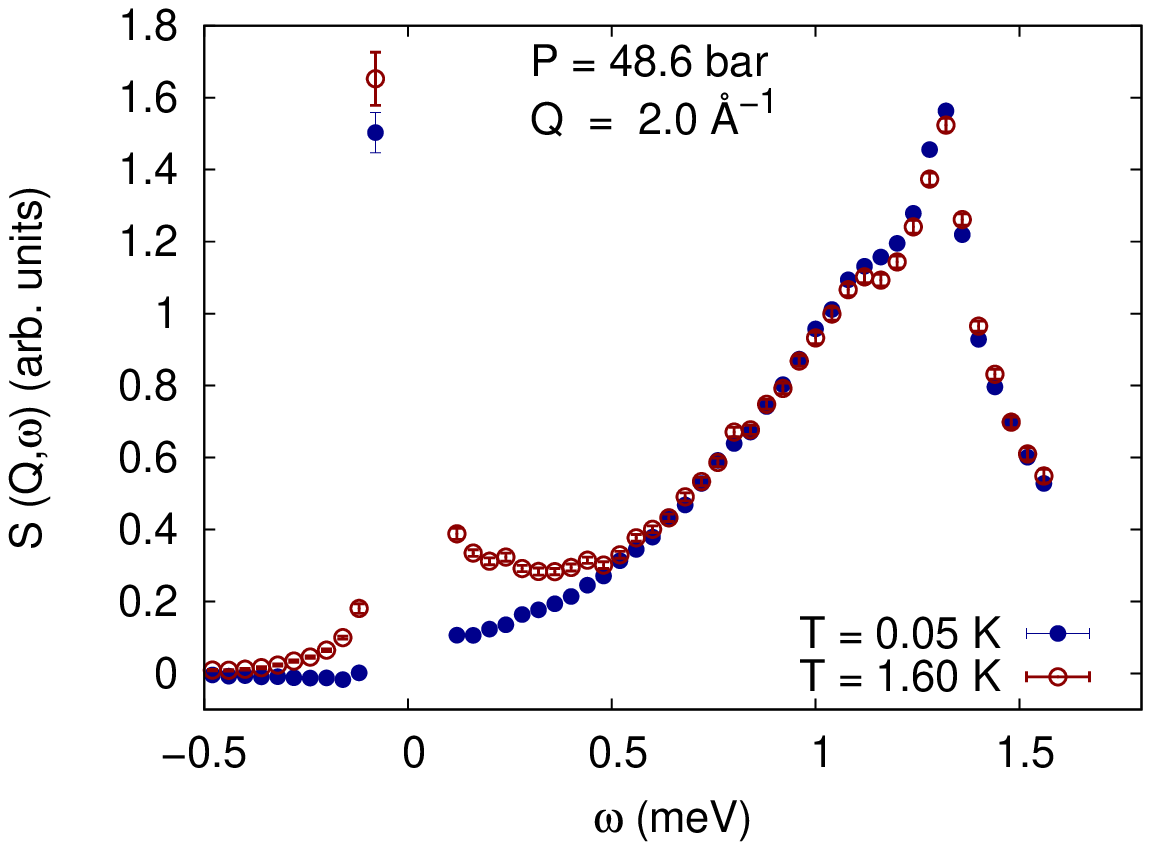}
\caption{
Net \sqw~ at Q = 2.0 \A of helium in MCM-41 at pressure $p$ = \pr~ at $T$ = \tlo~ (amorphous solid in the 
pores) and at $T$ = \thi~(liquid in the pores). \sqw~ includes the scattering from the bulk 
polycrystalline \4he lying between the grains of the MCM-41. At $T$ = \tlo, \sqw~ at low \w~ is 
approximately proportional to $\omega^2$ characteristic of scattering from a solid. \sqw~ has a peak at 
\w~$\simeq$ 1.3 meV attributed chiefly to the polycrystalline solid. At $T$ = \thi~ there is liquid in the 
pores with amorphous solid adjacent on the pore walls. \sqw~ at 1.6 K shows new intensity at low \w~ 
characteristic of normal liquid \4he.} 
\label{f2}
\end{figure}

Fig.~\ref{f2} shows the net DSF of helium in the present sample cell containing MCM-41 at pressure $p$ = 
48.6 bar. \sqw~at $Q$ = 2.0 \A~ and temperatures $T$ = 0.05 K and $T$ = 1.6 K is shown.  At $T$ = 0.05 K  
we expect amorphous solid helium in the MCM-41 pores and polycrystalline solid between the grains of the 
MCM-41 powder. The inelastic scattering (\w~$>$ 0) at $T$ = 0.05 K is a sum of \sqw~from the amorphous 
solid (30 \% of the helium sample in the beam) and polycrystalline solid (70 \% of the helium in the 
beam). At low $\omega$ the intensity in \sqw~grows  smoothly with \w~ as expected for scattering from a 
polycrystal or an amorphous solid in which the density of states g($\omega$) is approximately proportional 
to  $\omega^2$ at low \w.  There is no indication of any low energy roton modes or any layer modes as seen 
in liquid helium in porous media. The roton and layer modes of the liquid in MCM-41 are 
observed\cite{Pearce:04a,Bossy:08,Bossy:08a,Bossy:12} at energies 0.5 - 0.6 meV and 0.4 - 0.5 meV at 34 
bar, respectively. The DSF in Fig.~\ref{f2} peaks at $\omega\simeq 1.3$ meV. This is consistent with the 
peak in the density of states (DOS) of phonons in bulk polycrystalline helium between the grains observed 
previously.\cite{Bossy:08,Bossy:08a,Bossy:12} The large peak at $\omega = 0$ is elastic scattering, 
$S(Q,\omega = 0)$, from the amorphous solid in the pores. The elastic scattering from the polycrystalline 
solid is confined to Bragg peaks and the {\bf Q} in Fig.~\ref{f2} has been selected to avoid these Bragg 
peaks.

\begin{figure}[h]
\hspace{0.05cm}
\includegraphics[angle=0,width=0.5\textwidth]{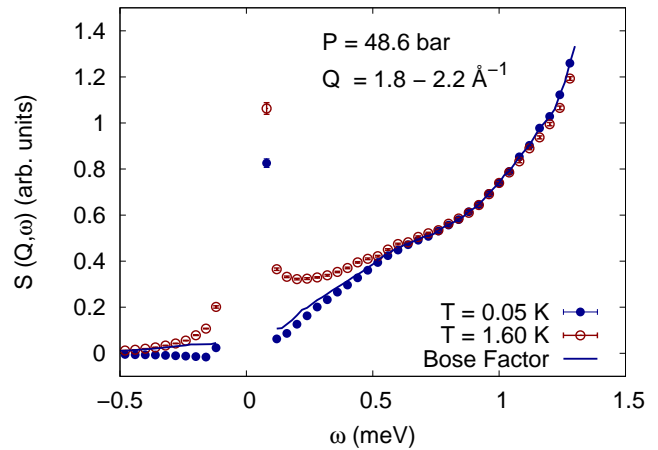}
\caption{
The net \sqw~ as in Fig. \ref{f2} integrated over $Q$ between $Q_1$ = 1.8 \A and $Q_2$ = 2.2 \A as defined 
in Eq. (1). The line labeled ``Bose Factor" is the data at T = \tlo~ multiplied by the Bose factor 
[$n_B(\omega)$ + 1] for temperature T = 1.6 K where $n_B(\omega)$ is the Bose function. The line displays 
the change in \sqw~ between 0.05 K and 1.6 K expected from thermal contributions alone. At $T$ = 1.6 K 
thermal contributions to \sqw~ are still clearly small.}
\label{f3}
\end{figure}

In Fig.~\ref{f2} at $T$ = 1.6 K there is new response at low $\omega$ in \sqw~not seen at $T$ = 0.05 K. 
Also at $T$ = 1.6 K the intensity in \sqw~at higher \w~(\w~$>$ 1 meV) is marginally smaller. At $Q$ = 2.0 
\A, energy transfers \w~ $>$ 1.5 meV are not observable on IN5 at the present incident neutron energy 
used.  From the phase diagram Fig.~\ref{f1}, we see that the helium in the interior of the MCM-41 pores is 
liquid at p = 48.6 bar and $T$ = 1.6 K.  The helium in the one-two helium layers tightly bound to the 
MCM-41 walls remains amorphous solid at $T$ = 1.6 K and higher temperatures.  Approximately 30 \% of the 
helium in the 47 \AA~MCM-41 pores is in the tightly bound layers adjacent to the pore walls which remains 
solid at $T$ = 1.6 K.  Thus approximately 20\% of the total helium sample in the beam melts to liquid 
between $T$ = 0.05 K and 1.6 K.  

The intensity in \sqw~at low \w~at $T$ = 1.6 K is attributed to the normal liquid \4he in the interior of 
the pores.  This \sqw~is similar to that of bulk normal liquid \4he at 20 bars\cite{Talbot:88,Gibbs:99} 
and of confined normal \4he in MCM-41 at 34 bars\cite{Bossy:12} as we discuss further below.  That is, 
\sqw~is concentrated near \w = 0 with a tail extending out to higher \w, as in normal 
liquids.\cite{Skold:72,Buyers:75}
Particularly, there are no sharp modes in \sqw~at either 0.05 K or 1.6 K characteristic of a 
Bose-condensed liquid.\cite{Bossy:12}

Fig.~\ref{f3} shows \sw~ obtained by integrating \sqw~ over a range of Q values,   
\begin{equation}
S(\omega) = \int\limits_{Q_1}^{Q_2} dQ S(Q, \omega),
\label{e1}
\end{equation}
between $Q_ 1$  = 1.8 \A~ and  $Q_ 2$  = 2.2 \A. The purpose of the integration is chiefly to improve the 
statistical precision of the data especially at low \w.  As in Fig.~\ref{f2}, the additional intensity at 
low \w~at $T$ = 1.6 K is attributed to normal liquid \4he in the interior of the pores.  The blue line in 
Fig.~\ref{f3} shows \sqw~of solid helium observed at T=0.05 K  (bulk polycrystalline helium between the 
grains and amorphous solid helium in the pores) multiplied by the Bose factor [$n_B(\omega)$ + 1]  where  
$n_B(\omega)$ is the Bose function for $T$ = 1.6 K  (0.15 meV).  The blue line represents the change in 
\sqw~of the solid expected when $T$ is increased from 0.05 K to 1.6 K if the energies and lifetimes of the 
modes of the solid remain unchanged.  Clearly, thermal effects make only a minor contribution to the 
observed change in \sqw~between $T$ = 0.05 K and $T$ = 1.6 K, even at low \w~for \w~$>$ 0.  Thus it is 
reasonable to attribute the observed large change in \sqw~at low \w~to melting of the amorphous solid in 
the interior of the pores to a liquid.  As in Fig.~\ref{f2}, no sharp or well defined modes are observed 
in the integrated \sqw.                          

\begin{figure}[h]
\hspace{0.05cm} 
\includegraphics[angle=0,width=0.5\textwidth]{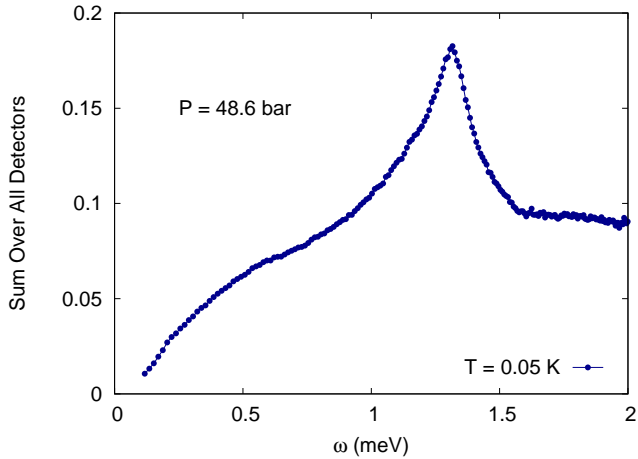} 
\caption{
The net \sw~ as defined in Eq. (\ref{e1}) and shown in Fig. (\ref{f3}) for amorphous solid in the pores 
and polycrystalline solid between the grains at $T$ = \tlo~ summed over all detectors. The sum over all 
detectors does not show any well-defined modes at low $E~\leq$~ 1 meV, only an \sw~ expected from a 
density of phonon like modes as given by the sum of  Eq. (\ref{d11}) describing a polycrystalline solid 
between the grains and Eq. (\ref{d18}) an amorphous solid in the pores.}
\label{f4}
\end{figure}

In Fig.~\ref{f3}, the intensity in the energy range  $0.6 < \omega < 1.5$ meV  is very similar at T = 0.05 
K and at T = 1.6 K. It is marginally smaller at 1.6 K than at 0.05 K but has a similar energy dependence. 
Thus the \sqw~ in this energy range arising from the amorphous solid ($T$ = 0.05 K) and the liquid ($T$ = 
1.6 K) in the interior of the pores appear similar but with marginally larger intensity from the amorphous 
solid. 

Fig.~\ref{f4} shows the \sw~ data arising from the amophous solid in the pores and the polyscrystalline 
solid between the grains at T = 0.05 K summed over all detectors.  The goal is to show that there are no 
sharply defined modes at \w~$\leq$ 1 meV at accessible $Q$ values. Rather \sw~ increases uniformly with 
\w~ as expected for scattering from phonons in a polycrystalline solid. The peak in the data summed over 
all detectors at 1.3 meV is broader than the peak in \sqw~ at $Q$ =2.0 \A shown in 
Fig. 1. The peak at 1.3 meV is expected to arise chiefly from the polycrystalline solid helium between the 
grains. Solid helium is a highly anharmonic solid so that there will be substantial anharmonic broadening 
in the observed \sqw~ and \sw.  In the Discussion we compare \sqw~ and \sw~ of anharmonic polycrystalline 
and amorphous solids and provide explicit expressions for the \sw~ observed from each.

\begin{figure}[h]
\hspace{0.05cm} 
\includegraphics[angle=0,width=0.5\textwidth]{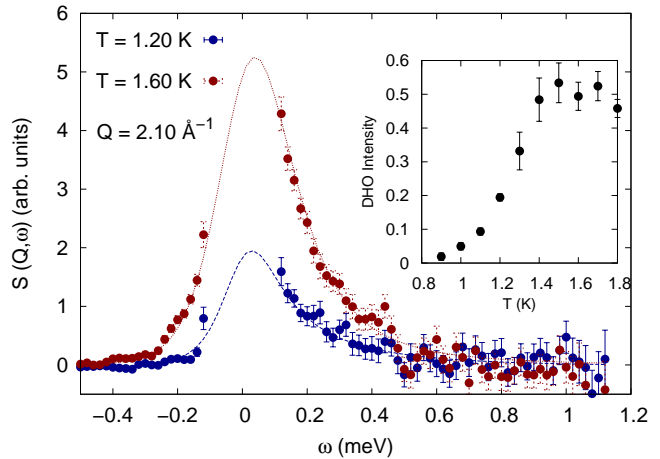} 
\caption{
The net \sqw~ of liquid \4he in MCM-41 at $T$ = 1.2 K and $T$ = 1.6 K interpreted as normal liquid. The 
lines are fits of a damped harmonic oscillator (DHO) function. The shape of \sqw~ is independent of 
temperature. Only the integrated intensity in \sqw~ increases with $T$ (see inset) as more amorphous solid 
melts to a liquid with increasing temperature. \sqw~ of normal liquid \4he at higher pressure peaks near 
\w~ = 0 as observed in classical liquids\cite{Buyers:75,Skold:72}.} 
\label{f5}
\end{figure}
%
\begin{figure}[h]
\hspace{0.05cm}
\includegraphics[angle=0,width=0.5\textwidth]{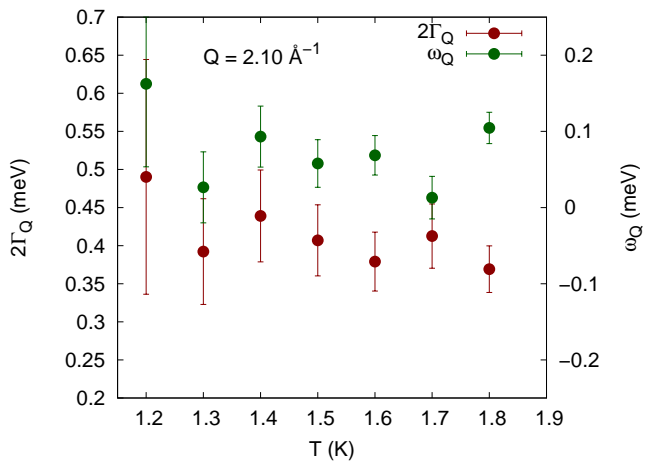} 
\caption{
The energy and width (FWHM) parameters of the DHO function that provides the best fit to the normal liquid 
\sqw~ shown in Fig. \ref{f4}. The parameters and therefore the shape of the normal liquid \sqw~ are 
independent of temperature. Only the magnitude of \sqw~ increases with increasing temperature.}
\label{f6}
\end{figure}  

Fig.~\ref{f5} shows the difference between \sqw~ with normal liquid \4he in the pores (at $T$ = 1.2 K and 
1.6 K) and amorphous solid in the pores at $T$ = 0.05 K, the latter multiplied by the thermal Bose factor 
for each temperature as discussed in Fig.~\ref{f5}. This difference is interpreted as the net scattering 
from the normal liquid at low \w. At higher \w, (\w$\geq$ 0.8 meV) the \sqw~ of the liquid and amorphous 
solid could be quite similar and the net \sqw~ is not well determined. The net liquid \sqw~ is a smooth 
function of \w~ as expected for normal \4he. The \sqw~ also peaks near \w~ = 0 as observed in classical 
liquids\cite{Buyers:75,Skold:72}. In contrast, the \sqw~ of normal liquid \4he at SVP ($p\approx$ 0) peaks 
at \w$\simeq$ 0.5 meV. Thus normal \4he at higher presssure responds much like a classical liquid. The 
lines in Fig.~\ref{f5} are fits of a damped harmonic oscillator (DHO) function, 
\begin{align}
 S_1(Q,\omega) = \frac{Z_Q/\pi}{1-\exp(-\hbar\omega/k_BT)}\nonumber\\
\times\left[\frac{4\omega\omega_Q\Gamma_Q}{{(\omega^2-[\omega_Q^2+\Gamma_Q^2])}^2+
4\omega^2\Gamma_Q^2}\right],
\label{e2}
\end{align}
to the data with energy, $\omega_Q$, width, $2\Gamma_Q$, and intensity, $Z_Q$, treated as free fitting 
parameters. The origin of the DHO is discussed in the appendices of Refs.\onlinecite{Talbot:88} and 
\onlinecite{Glyde:94} and the DHO is a standard fitting function in the literature.\cite{Gibbs:99} We 
found good fits to the data for energies and widths in the DHO function that were independent of 
temperature within precision (see Fig.~\ref{f6}). Only the weight or intensity in the DHO increased with 
increasing temperature between 0.8 K and 1.4 K. The weight (strictly the product $\omega_Q\Gamma_QZ_Q$ 
which fluctuates less than $Z_Q$ alone since $\omega_Q$ is nearly zero) is shown in the inset of 
Fig.~\ref{f5}. The increase in intensity with temperature was attributed to the increase in volume of 
liquid with temperature as the solid melts. The intensity in the liquid \sqw~ reaches a maximum at $T$ = 
1.4 K which is interpreted as the temperature when melting of the amorphous solid in the pores is 
complete. At lower pressure\cite{Bossy:12} we also found the energy and width of the normal liquid \sqw~ 
was independent of temperature over the temperature range investigated (1.2 $< T < 1.8$ K). No physical 
meaning is attributed to the energy or width in terms of modes. The DHO is only a convenient 
representation of \sqw.    

\begin{figure}[h]
\hspace{0.05cm} 
\includegraphics[angle=0,width=0.5\textwidth]{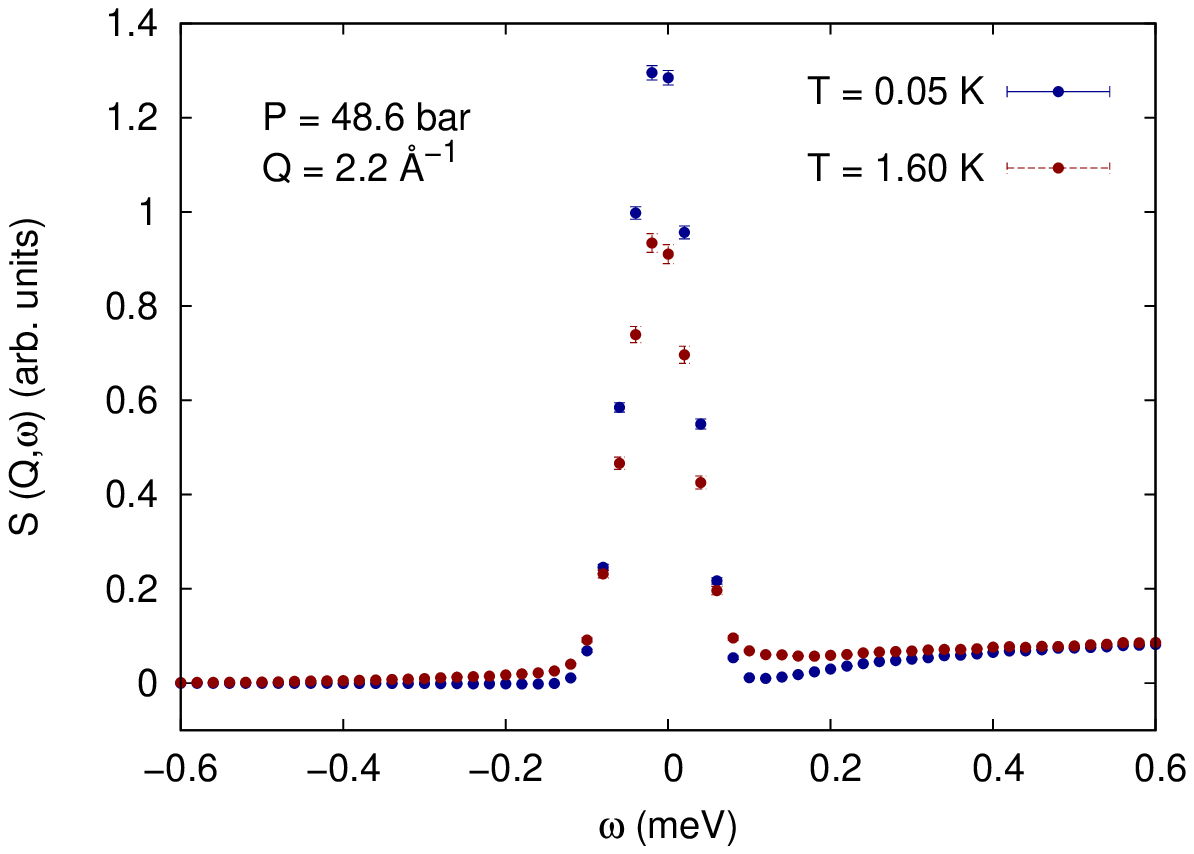} 
\includegraphics[angle=0,width=0.5\textwidth]{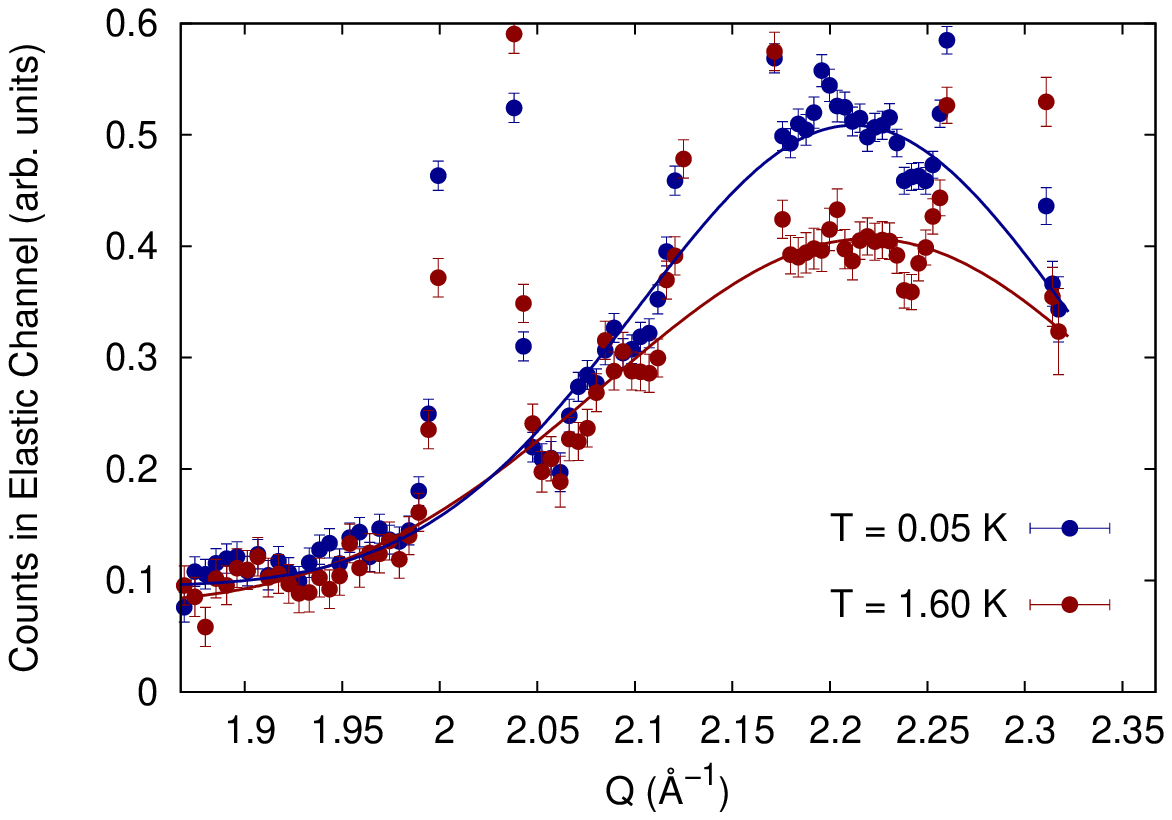} 
\caption{
The net elastic scattering from helium in and between the grains of MCM-41 at $T$ = 0.05 K and $T$ = 1.6 
K. At 0.05 K there is amorphous solid throughout the pores, at 1.6 K on the pore walls only. Top: \sqw~ 
with elastic peak at \w~ = 0 attributed to the amorphous solid. Bottom: $S(Q,\omega = 0)$ vs Q with lines 
through the amorphous component. Bragg peaks from the bulk hcp polycrystalline solid between the grains 
can be seen in $S(Q,\omega = 0)$ at $Q$ = 2.02 \A~(1000), 2.15 \A~(0002) and 2.29 \A~(1011).}  
\label{f7}
\end{figure}

Fig.~\ref{f7} shows the net elastic scattering from the helium in the sample cell at $T$ = 0.05 K and 1.6 
K.  The top frame of Fig.~\ref{f7} shows \sqw~ at $Q$ = 2.2 \A with the elastic peak at \w~ = 0. The 
elastic peak in the top frame arises chiefly from the amorphous solid helium in the pores or on the grain 
surfaces. The elastic scattering from the polycrystalline solid between the grains is confined chiefly to 
Bragg peaks which are not seen at $Q$ = 2.2 \A. At $T$ = 0.05 K the pores are full of amorphous solid. At 
1.6 K there is amorphous helium in the first 1-2 layers on the pore walls only, approximately 30 \% of the 
\4he in the pores. The elastic peak is clearly larger at 0.05 K indicating more amorphous solid than at 
1.6 K. However, the peak is only approximately 20 \% larger rather than a factor of 2-3 as might be 
expected if only amorphous solid contributes to the peak.

The bottom frame of Fig.~\ref{f7} shows the elastic scattering \sqw~ at \w~ = 0 as a function of $Q$. In 
this case Bragg peaks arising from the polycrystalline solid lying between the grains are observed at $Q$ 
= 2.02 \A and  $Q$ = 2.28 \A. The blue and red lines in Fig.~\ref{f7} are guides to the eye through 
elastic scattering arising from the amorphous solid at $T$ = 0.05 K and 1.6 K, respectively. Again, the 
intensity from the amorphous solid is greater at 0.05 K when the pores are filled with amorphous solid 
than at 1.6 K when there is amorphous solid in 1-2 layers on the pore walls only. The difference in 
intensity is comparable to that shown in the top frame of Fig.~\ref{f7}.
 
\begin{figure}[h!]
\hspace{-0.05cm}
\begin{flushleft}
\includegraphics[angle=0,width=0.5\textwidth]{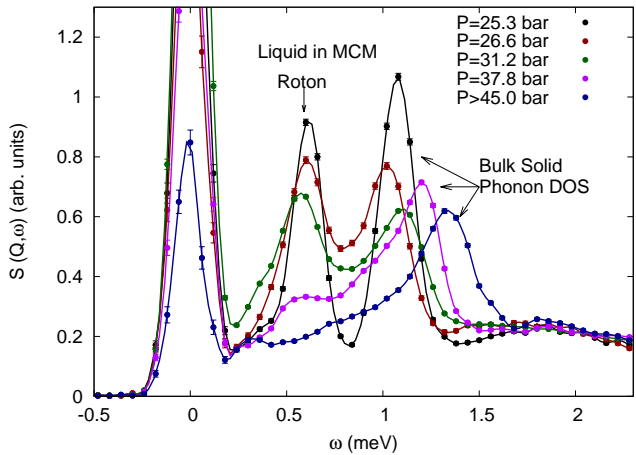} 
\end{flushleft}
\caption{\sqw~ at $Q$ = 2.1 \A~ (the roton $Q$) versus pressure at $T$ = 0.4 K from liquid \4he in MCM-41 
and bulk solid helium between the grains (from Ref. \onlinecite{Bossy:12}). The intensity arising from the 
roton of liquid \4he in MCM-41 decreases with increasing pressure until at 37.8 bar there is little or no 
roton. There is no roton at $p$ = 45 bar. The energy of the phonon DOS of the solid increases with 
increasing pressure.}  
\label{f8}
\end{figure} 
 
Finally, in Fig.~\ref{f8} we reproduce from Ref. \onlinecite{Bossy:12} measurements of \sqw~ at lower 
pressure where there is liquid \4he in the MCM-41 pores at low temperature, $T$ = 0.4 K. In Fig.~\ref{f8} 
we observe elastic scattering from the amorphous solid layers in the pores at \w~= 0, as in the top half 
of Fig.~\ref{f7} at $p$ = \pr. The peak at \w~= 0.6 meV is the roton mode of the Bose condensed liquid at 
$Q$ = 2.1 \A. The intensity in the roton mode clearly decreases with increasing pressure until there is 
only a very weak mode observed at $p$ = 37.8 bar which is immediately below the solidification pressure. 
No roton mode is observed at $p \simeq$ 45 bar. The decrease in intensity of the phonon-roton modes in the 
liquid with increasing pressure and disappearance of the mode at the solidification pressure $p$ $\simeq$ 
38 bar is consistent with no low-energy, liquid-like mode observed here in the amorphous solid at \pr. The 
broad peak at higher energy 
\w~ = 1.0 - 1.3 meV in Fig.~\ref{f8} is scattering from the phonons of the polycrystalline solid lying 
between the grains. The energy of the peak arising from the phonons increases with increasing pressure as 
expected for phonons. The broad peak at \w~$\simeq$ 1.2 meV at $p$ = 37.8 bar shown in Fig.~\ref{f8} is 
similar in shape to the peak in \sqw~ at \w~$\simeq$ 1.3 meV observed here at \pr, as seen in  
Figs.~\ref{f1} and \ref{f4}. In each case this peak is interpreted as arising chiefly from the phonons in 
the hcp polycrystalline solid between the grains.

\section{Discussion}

\subsection{Melting in confinement}

Taniguichi and Susuki\cite{Taniguchi:11} have recently reported measurements of freezing and melting of 
helium in a 28 \AA~ pore diameter FSM which is quite similar to the present MCM-41. They observe melting 
continuously over a wide temperature range, quite different from melting of bulk helium, but freezing over 
a narrower temperature range. In Fig. 5 above we presented the DSF, \sqw, of normal liquid \4he~ in the 
present MCM-41. Particularly, on warming, intensity in the liquid \sqw~ is first observed at T= 0.9 K, the 
intensity grows continuously until 1.5 K and saturates to a constant value at 1.5 K  (see inset of Fig. 
5). We interpret this as onset of melting at 0.9 K and melting complete at 1.5 K. This increase in 
intensity of the liquid \sqw~ over a wide temperature range is consistent with the continuous melting over 
a wide temperature range observed by Taniguchi and Susuki. They also observe a very small volume change on 
freezing, two orders of magnitude below the bulk value. This suggests that the molar volumes of the 
amorphous solid and liquid in FSM, and probably in the present MCM-41, are nearly identical. 
  
\subsection{Modes in confined liquid \4he}

In earlier measurements\cite{Pearce:04a,Bossy:08,Bossy:08a,Bossy:12} we have observed the DSF, \sqw, of 
liquid \4he in the present MCM-41 and in 25 \AA~diameter gelsil as a function of pressure and temperature. 
Both phonon-roton and layer modes are observed at low temperature.  The intensity in the roton decreases 
with increasing pressure, as noted above, and with increasing temperature. At $p$ = 34 bar, for example, 
the intensity in the roton disappears at $T$ = 1.5 K and a roton is no longer observed above this 
temperature. The solid circles, squares and triangles in Fig.~\ref{f1} show the maximum temperatures and 
pressures at which P-R modes are observed. At temperatures and pressures above the dashed black line 
through the data points in Fig.~\ref{f1}, well-defined P-R modes are no longer observed. Since well 
defined P-R modes exist when there is Bose-Einstein condensation, the dashed black line is associated with 
the temperature $T_{BEC}$ at which BEC disappears in liquid \4he confined in these porous media. In 25 
\AA~gelsil $T_{BEC}$ lies above $T_C$ for superflow as observed in a torsional 
oscillator.\cite{Yamamoto:04} For this reason, the temperature region $T_C < T < T_{BEC}$ is identified as 
a region of localized BEC with no superflow across the sample as discussed at the beginning of section 
III.

Equally interesting, the roton energy decreases with increasing pressure, decreasing from \dr~= 0.74 meV 
at SVP to approximately \dr~= 0.55 meV at 37.8 bar\cite{Pearce:04a,Bossy:08a}. The energy of a single P-R 
mode cannot exceed twice the roton energy\cite{Pitaevskii:59,Glyde:98}, 2\dr. If the single P-R mode 
energy exceeeds 2\dr, it has sufficient energy to spontaneously decay to two rotons and the mode will be 
broadened and not observable as well-defined mode. Thus
at higher pressure a well-defined liquid P-R mode exists at low energy only, at wave vectors in the phonon 
region and in the roton region only. For example, at $p \geq$ 38 bar (at solid pressures) a well-defined 
P-R mode exists in the liquid at energies $2\Delta \leq$ 1.1 meV only, i.e. at wave vectors in the phonon 
region $Q \leq$ 0.6 meV and in the roton region 1.7 $< Q <$ 2.4 \A only. Thus, even if there were trapped 
liquid in a solid at higher pressure, we would not observe a complete liquid P-R mode.  

In summary, our previous finding that the intensity in the liquid roton mode decreases with increasing 
pressure and goes to zero at $p\simeq$ 38 bar is consistent with the absence of a mode at roton energies 
in the amorphous solid in the present measurements.

\subsection{Amorphous Solid Helium}

In this section we compare expressions for the dynamic structure factor (DSF), \sqw, of an anharmonic 
polycrystalline and an anharmonic amorphous solid. The goal is to show that we expect the $S(Q, \omega)$ 
of solid helium in these two structures to be very similar when \sqw~ is integrated over a range of $Q$ 
values as in Eq.~(\ref{e1}) . The integrated \sw~ may be somewhat more sharply peaked as a function of 
$\omega$ in a polycrystalline solid because of the coupling between the scattering wave vector $Q$ and the 
wave vector, $q$, of the phonons in the polycrystalline solid. This coupling limits the number of phonons 
that can contribute to \sw~ in a polycrystal. However, because of (1) anharmonic effects which broadening 
the phonons substantially, (2) the large vibrational displacements of the atoms which make multimode 
contributions to $S(Q, \omega)$ large, and (3) because we are considering polycrystals rather than single 
crystals, we expect the difference to be small. To illustrate, we write out the single mode excitation 
term, $S_1(Q, \omega)$, of $S(Q, \omega)$ for polycrystalline and amorphous solids below. Higher order 
terms are discussed in an Appendix.

The total $S(Q,\omega)$ is the sum of an elastic scattering term, $S_0(Q)$, and a series of inelastic 
terms representing excitation of single modes, two modes, interference between those modes via anharmonic 
terms, and higher order mode processes,
\begin{align}
S(Q,\omega) = S_0(Q,0) + S_1(Q,\omega) + S_{INT}(Q,\omega) + S_2(Q,\omega) + ... \label{d1}
\end{align}
In a crystal of N atoms where there is periodic translational symmetry, we express the vibrational 
displacements $u_l(t)$ of the atoms $l$ from their lattice points $R_l$ as a superposition of waves 
(phonons) in the crystal of well defined wave vector $q$ and polarization index $\lambda$. With this 
expression, the product $[Q.u_l(t)]$ that appears in the DSF (see Eqs. (\ref{a2}) and (\ref{a6}) in the 
appendix) is, for one atom per unit cell,
\begin{align}
[Q.u_l(t)] = \frac{1}{\sqrt{N}} \sum_{q\lambda} f_{q\lambda} e^{iq\cdot R_l} [Q\cdot\epsilon_{q\lambda}] 
A_{q\lambda}(t).\label{d2}
\end{align}
In Eq. (\ref{d2}) $f_{q\lambda} = (\hbar/2m\omega_{q\lambda})^{1/2}$ where $\omega_{q\lambda}$ is the 
frequency and $\epsilon_{q\lambda}$  is the polarization vector of the wave and $m$ is the atomic mass. 
$A_{q\lambda}(t)$ is the mode annihilation operator for the phonon $(q=1$ to $N$ and $\lambda=1$ to 3). 

In randomly oriented polycrystals of solid helium, we assume that, averaged over the polycrystals,
\begin{align}
\langle [Q.\epsilon_{q\lambda} ]^2 \rangle_{poly} = \frac{1}{3}Q^2
\label{d3}
\end{align}
We restrict ourselves here to positive $\omega$ and low temperature $\hbar\omega >> kT$ so that the number 
of the thermally excited phonons is small. In this case, using the substitution (\ref{d2}) and the average 
(\ref{d3}), the one-phonon component of $S(Q,\omega)$ is 
\begin{align}
S_1(Q,\omega) = I_1(Q) \frac{1}{3N} \sum_{q\lambda} \frac{A(q\lambda,\omega) N\Delta (Q - q - 
\tau)}{2\pi\omega_{q\lambda}}  \label{d4}
\end{align}
%
in which $I_1(Q)$ is the one-phonon weight factor, $I_1(Q) = d^2(Q)(\hbar Q^2/2m)$, $d^2(Q)$ is the 
Debye-Waller factor and $A(q\lambda, \omega)$ is the one-phonon response function. In an anharmonic 
crystal such as solid helium, $A(q\lambda, \omega)$ has the form\cite{Cowley:68,Glyde:94},
\begin{align}
A(q\lambda, \omega) &= \frac{8\omega_{q\lambda}^{2} \Gamma(q\lambda, \omega)}{[-\omega^2 + 
\omega_{q\lambda}^{2} + 2\omega_{q\lambda}\Delta]^2 + [2\omega_{q\lambda} \Gamma]^2},
\label{d5}
\end{align}
where $\Gamma$ = $\Gamma(q\lambda,\omega)$ is the half width of the phonon group and $\Delta$ = 
$\Delta(q\lambda,\omega)$ is a shift in the initial frequency $\omega_{q\lambda}$ arising from anharmonic 
effects. In solid helium, the $\omega_{q\lambda}$ are typically some positive anharmonic frequencies such 
as self-consistent harmonic frequencies. The  $\Gamma(q\lambda, \omega)$ is large so that $A(q\lambda, 
\omega)$ is a broad function of $\omega$, especially for phonons near the Brillouin zone 
edges\cite{Minkiewicz:73,Markovich:02}.
In a harmonic limit in which $\Gamma \rightarrow 0$, the $A(q\lambda, \omega)$ reduces to
\begin{align}
A(\lambda,q\omega) = 2\pi \left[ \delta(\omega - \omega_{q\lambda}) -  \delta(\omega - 
\omega_{q\lambda})\right].\label{d6}
\end{align}
For positive energies, $\omega>0$, only the first term in $A(q\lambda, \omega)$ in Eq.~(\ref{d6}) 
contributes to $S_1(Q, \omega)$.

At $p$=\pr~ the one-phonon weight factor $I_1(Q) = d^2(Q)(\hbar Q^2/2m)$ has a broad peak at $Q\simeq 
2.3~\mathring{A}^{-1}$. Thus for polycrystalline solid helium, we expect $S_1(Q, \omega)$ to be largest at 
$Q$ values where $I_1(Q)$ has its maximum and at energies where $\omega_{q\lambda}$ is small.

The phonon density of states (DOS) of a harmonic crystal is defined as
\begin{align}
g(\omega) = \frac{1}{3N} \sum_{q\lambda} \delta(\omega - \omega_{q\lambda}) \label{d7}
\end{align}
where the sum is over one complete Brillouin zone. The $S_1(Q, \omega)$ in Eq. (\ref{d5}) is proportional 
to a modified DOS
\begin{align}
\tilde{g}_C(\omega) &= \frac{1}{3N} \sum_{q\lambda} \frac{1}{2\pi\omega_{q\lambda}} A(q\lambda,\omega) 
\label{d8}\\ 
&\simeq \frac{1}{3N} \sum_{q\lambda} \frac{1}{\omega_{q\lambda}} \delta(\omega - 
\omega_{q\lambda})\label{d9}
\end{align}
which is the usual DOS weighted by $\omega_{q\lambda}^{-1}$. The $\tilde{g}_C(\omega)$ is the modified DOS 
that is always observed in neutron scattering measurements\cite{Lovesey:84}. Eq. (\ref{d8}) is the general 
anharmonic DOS that will be observed in \sqw~ of solid helium. The second expression, Eq. (\ref{d9}), for 
$\tilde{g}_C(\omega)$ holds only in a harmonic approximation. Finally, if we integrate $S_1(Q, \omega)$ 
over a range of $Q$ values as in Eq. (\ref{e1}), 
we obtain
\begin{align}
S_1(\omega) &= \int dQ S_1(Q, \omega)\label{d10}\\
&=\left[ (2\pi)^3/\Omega\right] \int dQ I_1(Q) \tilde{g}_C(\omega) \delta(Q - q - \tau) \label{d11}
\end{align}
%
where we have used $N\Delta(Q)=[(2\pi)^3/\Omega] \delta(Q)$ and $(\Omega=V/N)$ is the volume of the unit 
cell. In Eqs. (\ref{d4}) and (\ref{d11}), there is a coupling between $Q$ and the phonon wave vector $q$ 
which means that certain values of $q$ only will contribute to $S_1(\omega)$. A full DOS is not observed. 
In solid helium $A(q\lambda, \omega)$ is a broad function in $\omega$, so that only an anharmonic DOS 
given by Eq. (\ref{d8}), substantially broadened by anharmonic terms, is observed.  In Fig.~\ref{f4} we 
show an \sw~ that is integrated over a limited range of $Q$ only. Thus we expect the corresponding 
$S_1(\omega)$ to be more sharply localized in $\omega$ (representing selected phonons) than the full 
modified DOS $\tilde{g}_C(\omega)$ given by (\ref{d8}).

In an amorphous solid, the mean positions $R_l$ do not have periodic symmetry. We treat the amorphous 
solid as a large molecule or as a solid that has a single large unit cell with all $N$ atoms in the unit 
cell\cite{Carpenter:85}. We express $u_l(t)$ as a superposition of the normal modes of the molecule 
(numerated by $\lambda$, $\lambda=1$ to $3N$) so that Eq. (\ref{d2}) becomes,
\begin{align}
[Q\cdot u_l (t)] &= \frac{1}{\sqrt{N}} \sum_{\lambda} f_{\lambda} \left[ Q\cdot\epsilon_{\lambda l} 
\right] A_{\lambda} (t) \label{d12}
\end{align}
In this case the polarization vectors $\epsilon_{\lambda l}$ have an index $l$. We assume that averaged 
over the $N$ atoms in the amorphous solid the polarization vector fulfil,
\begin{align}
\langle[Q\cdot\epsilon_{\lambda l}]\rangle_{Amor} &= \frac{Q}{\sqrt{3}}. \label{d13}
\end{align}
With this assumption, and restricting ourselves to positive \w~ and low temperatures  $\hbar\omega >> kT$ 
as before, $S_1(Q, \omega)$ for an amorphous solid reduces to,
\begin{align}
\nonumber S_1(Q,\omega) &= S_M(Q) I_1(Q) \frac{1}{3N} \sum_{\lambda} \frac{1}{2\pi\omega_{\lambda}} 
A(\lambda, \omega)\\
 & = S_M(Q) I_1(Q)  \tilde{g}_A(\omega) \label{d14}
\end{align}
where $S_M(Q) = {1/N} \sum_{l,l^{'}} \exp(-iQ\cdot[R_l-R_{l^{'}}])$ is a static structure factor defined 
in terms of the mean positions $R_l$ of the atoms, $I_1(Q) = d^2(Q)(\hbar Q^2/2m)$ is the one mode 
intensity as before, $A(\lambda, \omega)$ is the one mode response function of the form (\ref{d5}) and %
\begin{align}
\tilde{g}_A(\omega) &= \frac{1}{3N} \sum_{\lambda} \frac{1}{2\pi\omega_{\lambda}} A(\lambda, 
\omega)\label{d15}\\
 & \simeq \frac{1}{3N} \sum_{\lambda} \frac{1}{\omega_{\lambda}} \delta(\omega - \omega_{\lambda}).
\label{d16}
\end{align}
is a modified DOS for the amorphous solid. In a HA, $A(\lambda, \omega)$ reduces to a $\delta$ function 
and Eq. (\ref{d16}) becomes exact. The corresponding  $S_1(\omega)$ intergrated over a range of $Q$ values 
is
\begin{align}
S_1(\omega) &= \int dQ S_1(Q, \omega)\label{d17}\\
&= \tilde{g}_A(\omega) \int dQ S_M(Q) I_1(Q). \label{d18}
\end{align}
There are two important differences between $S_1(\omega)$ for a polycrystal, Eq. (\ref{d11}), and 
$S_1(\omega)$ for an amorphous solid, Eq. (\ref{d18}). In the amorphous solid we observe the modified DOS 
$g_A(\omega)$ directly in $S_1(Q, \omega)$ unaffected by selection of particular modes via a delta 
function. Also, in the amorphous solid $S_1(Q, \omega)$ contains an additional factor of $S_M(Q)$, which 
as $I_1(Q)$, peaks at $Q\simeq 2.3~\mathring{A}^{-1}$, in solid \4he at $p$ = \pr. Thus, we expect $S_1(Q, 
\omega)$ in the amorphous solid to be somewhat more sharply peaked in $Q$ (at $Q\simeq 
2.3~\mathring{A}^{-1}$) than in the polycrystal. If  $\tilde{g}_C(\omega)$ and  $\tilde{g}_A(\omega)$ are 
similar, we expect the $S_1(\omega)$ for a polycrystal to be somewhat more sharply peaked in $\omega$ 
because of the delta function selection of $q$ values in the polycrystalline case. However, given that 
$A(\lambda, \omega)$ is itself a broad function of \w, we expect the difference to be small.

In addition to the single mode $S_1(Q, \omega)$ there are higher mode terms $S_2(Q, \omega)$, $S_3(Q, 
\omega)$ in (\ref{d1}) which are not negligible in solid helium. The $S_2(Q, \omega)$, which is discussed 
in the appendix, is proportional to $I_2(Q) = d^2(Q)(\hbar Q^2/2m)^2$ and has a broad maximum at somewhat 
higher $Q$ values than $S_1(Q, \omega)$. It peaks in $\omega$ at higher $\omega$ than $S_1(Q, \omega)$. 
The chief effect of the higher order terms is to further broaden $S(Q, \omega)$ and to extend \sqw~ to 
higher $\omega$. 

Also, quite generally, at higher \w~ we expect the $S(Q, \omega)$ to be independent of phase (liquid or 
solid) and to become quite similar in polycrystalline, amorphous and liquid helium. Ultimately the 
response at high $\omega$ arises from high energy interaction between pairs of atoms via the hard core of 
the interatomic partial which will be independent of structure and similar in all three phases.
This high energy region begins at energies greater than the collective mode (e.g. phonon) energies. Thus 
we expect \sw~ of amorphous and crystalline \4he to be similar at high \w~ as well as at lower \w~ from 
the arguments above. 

\section{Conclusion}

The dynamical response of amorphous solid helium confined in MCM-41 at 48.6 bar as observed in the 
dynamical structure factor, \sqw, is a smooth function of energy (\w) characteristic of a solid that has a 
vibrational density of states approximately proportional to \w$^2$ at low \w. No sharp excitation at low 
\w~ similar to the phonon-roton mode in Bose-condensed liquid helium is observed. The \sqw~ of amorphous 
and bulk polycrystalline solid helium are similar and broad at $Q$ values around 2 \A, as anticipated for 
a highly anharmonic solid.  Above $T$ = 1 K, the amorphous solid melts to normal liquid \4he. The \sqw~ at 
$Q\simeq$ 2 \A~ of the normal liquid at 48.6 bar is a broad function that peaks near \w~$\simeq$ 0 as in 
classical liquids, rather than a broad function peaking at a finite \w~ (e.g. \w~= 0.5 meV) as in (more 
quantum) normal liquid \4he at $p\simeq$ 0.

\section{Acknowledgement}

It is a pleasure to acknowledge the support of the Institut Laue Langevin and O. Losserand and X. Tonon at 
ILL for valuable assistance with the experiments. This work was supported by the DOE, Office of Basic 
Energy Sciences under contract No. ER46680.

\section{Appendix}
In this appendix, we present some background on the dynamic structure factor, $S(Q, \omega)$, to support 
the expressions given in section IV B. We begin with the coherent, intermediate 
scattering function,
\begin{align}
S(Q, t) &= \frac{1}{N} \sum_{l,l^{'}} \langle e^{-iQ\cdot r_l(t)} e^{iQ\cdot r_{l^{'}}(0)}\rangle 
\label{a1}\\
& \approx d^2(Q) \frac{1}{N} \sum_{l,l^{'}} e^{-iQ\cdot[R_l-R_{l^{'}}]} e^{\langle[Q\cdot u_l(t)][Q\cdot 
u_{l^{'}}(0)]\rangle}, \label{a2}
\end{align}
in which $r_l(t) = R_l + u_l(t)$ where $R_l$ is the mean position of atom $l$ (lattice vector in a 
crystal) and $u_l(t)$ is the displacement of the atom from $R_l$ at time $t$. $Q$ is the wave vector 
transfer in the scattering and $d^2(Q)$ is the Debye-Waller factor, $d^2(Q) = \exp\left[-Q^2 \langle u^2 
\rangle / 3\right]$ in a cubic crystal. We consider  one atom per unit cell and cubic symmetry for 
simplicity. Eq.~(\ref{a2}) omits the anharmonic interference terms. The corresponding DSF is
\begin{align}
S(Q,\omega) = \frac{1}{2\pi} \int dt e^{i\omega t} S(Q,t). \label{a3}
\end{align}

To generate the series Eq. (\ref{d1}), we expand the second exponential in Eq. (\ref{a2}) in a power 
series in $\langle[Q.u_l(t)][Q.u_{l^{'}}(0)]\rangle^n $ in the usual way\cite{Lovesey:84,Glyde:94}. The 
zero order term $(n=0)$ is the elastic term. For a crystal in which the $R_l$ has periodic order, the 
corresponding elastic DSF is 
\begin{align}
\nonumber S_0(Q,0) &= d^2(Q) \frac{1}{N} \sum_{l,l^{'}} e^{-iQ\cdot[R_l-R_{l^{'}}]}\delta(\omega) \\
&= d^2(Q)N\Delta(Q -\tau)\delta(\omega),\label{a4}
\end{align}
The intensity in $S_0(Q,0)$ is confined to Bragg peaks at the reciprocal lattice vectors $\tau$. In the 
present measurements involving bulk polycrystalline solid lying between the grains of the MCM-41, we 
generally chose wave vectors $\bf{Q}$ to avoid the Bragg peaks. In this way, elastic scattering from the 
polycrystallic solid is not observed. Above 1K this is not always easy since above 1K the polycrystals are 
continually re-crystallizing\cite{Burns:08,Bossy:10} and a Bragg peak may appear during a measurement.

For an amorphous solid, where the mean positions $R_l$ of the atoms do not have periodic order, 
\begin{align}
\nonumber S_0(Q,0) &= d^2(Q) \frac{1}{N} \sum_{l,l^{'}} e^{-iQ\cdot(R_l-R_{l^{'}})} \delta(\omega)\\
& = d^2(Q)S_M(Q)\delta(\omega)\label{a5}
\end{align}
where $S_M(Q)$ is a static structure factor defined by the mean positions, $R_l$ of the atoms. The elastic 
$S_M(Q)$ is not the same as the static structure factor $S(Q)$. We have found that the $S(Q)$ of the 
amorphous solid is somewhat more sharply peaked in the peak region than that of the liquid but otherwise 
the $S(Q))$ of the liquid and amorphous solid are very similar.\cite{Bossy:10}

The inelastic scattering in which the neutron creates or annihilates a single mode arises from the term 
proportional to $\langle[Q.u_l(t)][Q.u_{l^{'}}(0)]\rangle$ $(n=1)$ in the expansion of Eq.~(\ref{a2}). The 
corresponding intermediate DSF is,
\begin{align}
S_1(Q, t) = d^2(Q) \frac{1}{N} \sum_{l,l^{'}} e^{iQ\cdot[R_l - R_{l^{'}}]} \langle[Q\cdot u_l(t)][Q\cdot 
u_{l^{'}}(0)]\rangle . \label{a6}
\end{align}
For a crystal, we expand $[Q.u_l]$ in Eq.~(\ref{a6}) in terms of phonon modes as in Eq.~(\ref{d2}). The 
single mode term of $S(Q, \omega)$ for a crystal is then, 
\begin{align}
\nonumber S_1(Q,\omega) = &\frac{1}{2\pi} \left[ n_B(\omega) +1\right]  d^2(Q) \frac{1}{N} \sum_{q\lambda} 
f^{2}_{q\lambda} \left[ Q\cdot\epsilon_{q\lambda} \right]^2 \\
& \times A(q\lambda,\omega) N\Delta (Q - q - \tau). \label{a7}
\end{align}
where $A(q\lambda, \omega)$ is the one phonon response function given by Eq.~(\ref{d5}) in general and by 
Eq.~(\ref{d6}) for a harmonic crystal. Making the assumption Eq.~(\ref{d3}) for a polycrystal and in the 
limit of low temperatures where $n_B(\omega)$ may be neglected, Eq.~(\ref{a7}) reduces to Eq.~(\ref{d7}). 

For an amorphous solid, we expand the $[Q.u_l]$ in terms of the modes $\lambda$ of the molecule given by 
Eq.~(\ref{d12}). The resulting $S_1(Q,\omega)$ for an amorphous solid is, 
\begin{align}
\nonumber S_1(Q,\omega) =& \frac{1}{2\pi} [n_B(\omega) + 1] d^2(Q) \sum_{l,l^{'}} 
e^{iQ\cdot(R_l-R_{l^{'}})} \\
& \times \frac{1}{N} \sum_{\lambda} f_{\lambda}^{2} [Q\cdot\epsilon_{\lambda l}][Q\cdot\epsilon_{\lambda 
l^{'}}] A(\lambda, \omega). \label{a8}
\end{align}
With the assumption Eq.~(\ref{d14}) for the $[Q.\epsilon_{\lambda l}]$ and neglecting $n_B(\omega)$, Eq.~ 
(\ref{a8}) reduces to Eq.~(\ref{d15}) for amorphous solids. As discussed in section IV B, the coherent 
$S_1(Q, \omega)$ of an amorphous solid is directly proportional to the DOS $\tilde{g}_A(\omega)$ that 
would be observed in the incoherent DSF. There is no well defined wave vector $q$ and therefore no 
coupling between $q$ and $Q$ as there is in the crystalline case. For an anharmonic solid, we expect 
$S_1(\omega)$ in the crystalline and amorphous phases to both be substantially broadened by anharmonic 
effects which will make the two quite similar.

The two phonon term $S_2(Q,\omega)$ is the term proportional to 
$\langle[Q.u_{l^{'}}(t)][Q.u_{l^{'}}(0)]\rangle^2$ in the expansion of  Eq.(\ref{a2}). Substituting the 
expansion Eq.~(\ref{d2}) in this term, we obtain, 
\begin{align}
\nonumber S_2(Q,\omega) =& d^2(Q) \frac{1}{2\pi} [n_B(\omega)+1] \frac{1}{2N}\sum_{q_1,\lambda} 
\sum_{q_2,\lambda} f_{2}^{2} f_{1}^{2}\\
 & \times [Q\cdot\epsilon_1]^2 [Q\cdot\epsilon_2]^2 \Delta (Q - q_1 - q_2 - \tau )  A_2(12, \omega) 
\label{a9}
\end{align}
where 
\begin{align}
\nonumber [n_B(\omega)+1] A_2(12, \omega) =& \int_{-\infty}^{\infty} \frac{d\omega^{'}}{2\pi} 
[n_B(\omega^{'})+1] A(1, \omega^{'})\\
& \times [n_B(\omega - \omega^{'})+1] A(2, \omega - \omega^{'})  
\end{align}
\noindent and where $1=q_1\lambda_1$ and $2=q_2\lambda_2$. For a polycrystalline sample in which Eq. 
(\ref{d3}) is assumed to hold and in which the temperature is low, $S_2(Q, \omega)$ reduces to 
\begin{align}
\nonumber S_2(Q, \omega) =& d^2(Q) \left(\frac{\hbar Q^2}{2M} \right)^2 \frac{1}{2} 
\left(\frac{1}{3N}\right)^2 \sum_{q_1\lambda_1 q_2\lambda_2} \frac{1}{2\pi\omega_1 \omega_2}\\
& \times A(12, \omega) N\Delta(Q - q_1 - q_2 - \tau) \label{a10}
\end{align}
where
\begin{align}
A(12, \omega) & = \int \frac{d\omega^{'}}{2\pi} A(1, \omega^{'}) A(2, \omega - \omega^{'}) \label{a11}
\end{align}
Since $A(q\lambda, \omega)$ is a broad function, the $A(12, \omega)$  will be an even broader function 
since $A(12, \omega)$ is a convolution. $S_2(Q, \omega)$ is significant in solid helium and contributes to 
$S(Q,\omega)$ at higher $\omega$. When $S_2(Q, \omega)$ is significant,  we expect $S(\omega)$ given by 
(\ref{e1}) to be even more similar for amorphous and crystalline solids, especially at higher $\omega$.

\bibliographystyle{apsrev}

\end{document}